\documentclass[english,aps,prl,amsmath,amssymb,letterpaper,showpacs,superscriptaddress, twocolumn, notitlepage, groupedaddress]{revtex4-1}
\usepackage{amsmath}
\usepackage{amssymb}
\usepackage{float}
\usepackage{graphicx}
\usepackage[usenames,dvipsnames]{color}
\usepackage{hyperref}
\usepackage{bbm}
\usepackage{pgfkeys}
\usepackage{url}

\usepackage{tikz}
\usetikzlibrary{calc, shapes, backgrounds, snakes}

\tikzset{
  mm/.style = {fill = yellow!90!yellow,
                 label = center:\textsf{mm}},
  mf/.style = {fill = blue!70!yellow, text = black,
                 label = center:\textsf{mf}},
  fm/.style = {fill = blue!70!yellow, text = black,
                 label = center:\textsf{fm}},
  ff/.style = {fill = yellow!90!yellow, text = black,
                 label = center:\textsf{ff}},
  croak/.style = {fill = green!90!yellow, text = black,
                 label = center:\textsf{c}},
  nocroak/.style = {fill = red!90!yellow, text = black,
                 label = center:\textsf{$\bar{\text{c}}$}},
  lcroak/.style = {fill = green!90!yellow, text = black,
                 label = center:\textsf{l}},
  rcroak/.style = {fill = red!90!yellow, text = black,
                 label = center:\textsf{r}}
}

\begin{document}

\title{When intuition fails in assessing conditional risks: the example of the frog riddle}

\author{Daniel Hetterich}
\author{Florian Gei\ss ler}
\affiliation{Institut f\"{u}r Theoretische Physik und Astrophysik,
Universit\"{a}t W\"urzburg, D-97074 W\"urzburg, Germany}
 \date{\today}

\begin{abstract}
Recently, the educational initiative TED-Ed \cite{youtube} has published a popular brain teaser coined the ``frog riddle'', which illustrates non-intuitive implications of conditional probabilities. 
In its intended form, the frog riddle is a reformulation of the classic boy-girl paradox \cite{Gardner54}. 
However, the authors alter the narrative of the riddle in a form, that subtly
changes the way information is conveyed.
The presented solution, unfortunately, does not take this point into full account, and as a consequence, lacks consistency in the sense that different parts of the problem are treated on unequal footing.
We here review, how the mechanism of receiving information matters, and why this is exactly the reason that such kind of problems challenge intuitive thinking.
Subsequently, we present a generalized solution, that accounts for the above difficulties, and preserves full logical consistency. Eventually, 
the relation to the boy-girl paradox is discussed.
\end{abstract}
\maketitle
%\begin{keywords}
%conditional probabilities, boy-girl paradox, problem solving
%\end{keywords}

\section{Introduction and definition of the problem}

%\paragraph{Description of the riddle:}
The brain teaser presented in the educational video \cite{youtube} is the following: You find yourself lost in the jungle, and even worse, 
poisoned by a mushroom. The only chance to survive is to catch one of the frogs living in the jungle, of which the \textit{female} specimen produce an antidote. On the other hand, while the \textit{male} frogs have a distinct croak, the females are mute, which makes them harder to identify. Female and male frogs cannot be distinguished from their appearance, and are equally numerous in the jungle.
Luckily, you spot a single frog sitting on a tree stump in front of you. At that very moment, you hear a single croak from behind. Turning around, you see two frogs in a clearing. Apparently, at least one of the two frogs (the one that just croaked) is male.
As time is running out, you quickly have to make a decision about where to go, in order to catch a female frog and get the antidote: should you go for the single frog, or for the two frogs, of which at least one is a male? What are your chances of survival in either direction? Importantly, we assume, that going to the clearing with two frogs implies, that you catch \textit{both} the frogs there.

%\paragraph{TED's presented solution:} 
Let us first discuss the solution presented by TED.
The authors of the riddle~\cite{youtube} consider it intuitive to assume, that the survival chance at the single frog, i.e. the chance that this frog is female (f = 'female'), is $P_\text{single}(\text{f})=1/2$. This value is inferred from the fact that male (m='male') and female frogs are found in equal numbers in the jungle. 
On the other hand, the chance to find a female among the two frogs is less intuitive, 
and TED suggests a survival chance of $P(\text{f}|\text{c})=P_\text{two}(\text{f}|\text{c})=2/3$, given that we heard a single croak (c='croak'). For clarity, we drop the label 'two' in the following whenever we refer to the two frogs, but denote the probability explicitly, when we refer to the single frog. In order to obtain the above survival chance of $2/3$ at the two frogs, the authors of Ref.~\cite{youtube} first consider the full sample space \textit{before} the croak.  
The two frogs could be mm = 'two males', ff = 'two females', and mf or fm, which are 'one male and one female'. The order is not important, but we choose to write mf and fm separately, such that all the four events mm, ff, mf, and fm are equally likely. In this notation, one may associate the label with the positions of the frogs, so mf = 'left frog male and right frog female'. 
As one of the two frogs croaks, this rules out the possibility of two females, because only males can croak. The equally likely events mm, fm, and fm remain, of which two out of three contain a female. Thus, the chance to find a female among both frogs is $P(\text{f}|\text{c})=2/3$.

\begin{figure}
\centering
\includegraphics[width=0.99\linewidth]{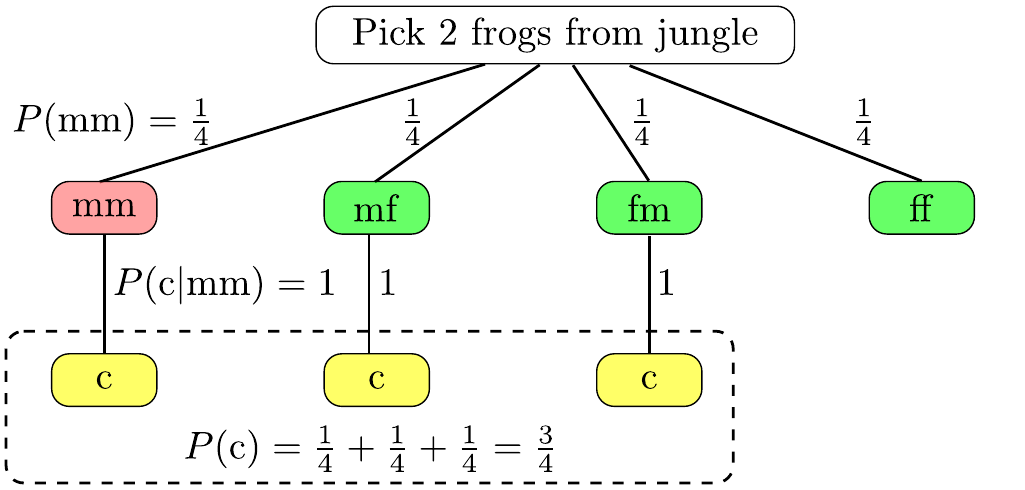}
\caption{(color online) Probability tree that follows the argumentation of Ref.~\cite{youtube}: If at least one male exists, there will be exactly one croak (c, yellow). 
From the three paths that lead to a croaking event, two entail the occurance of a female frog (f, green), resulting in a survival chance of $2/3$.}
\label{fig:tree0}
\end{figure}

%\paragraph{Problems with TED's solution:}
The above argumentation turns out to be problematic, and gives rise to
ambiguities, once the riddle is treated properly by means of conditional probabilities. We calculate the probability $P(\text{f}|\text{c})$, which is the probability of at least one female, \emph{given that} a croak was heard. As the opposite of 'at least one female' is 'no female', we find $P(\text{f}\vert \text{c}) = 1 - P(\text{mm}\vert \text{c})$. 
We can use Bayes' theorem to obtain
\begin{equation}
P(\text{f}\vert \text{c}) = 1- \frac{P(\text{c}\vert \text{mm}) P(\text{mm})}{P(\text{c})}=1-\frac{1 \cdot 1/4}{3/4}=2/3.
\label{eq:intended}
\end{equation}
We note, that unlike the chance $P(\text{mm})=1/4$, the two probabilities $P(\text{c}\vert \text{mm})$ and $P(\text{c})$ are not given by the narrative of the riddle.
In order to find the presented values, the authors of Ref.~\cite{youtube} \emph{implicitly} use a rule, which we could phrase as: 'A male frog always croaks, but if there are two males, there will also be only one croak.'
Only with this rule, one finds $P(\text{c}\vert \text{mm}) = P(\text{c}\vert \text{mf}) = P(\text{c}\vert \text{fm}) = 1$ and $P(\text{c}) = 3/4$. This point is illustrated in the tree diagram of Fig.~\ref{fig:tree0}.

Although the implicit rule mentioned above appears to be a natural approach to simplify the problem, it needs to be taken with care.
First of all, one may wonder the following. If the two frogs behave all the same -- meaning that we neglect any implications of two frogs sitting next to each other, such as that, for instance, a male frog croaks more often in the vicinity of a female -- why should two males croak as likely as a just one male? But more importantly, if males  always croak, the single frog on the tree trunk must be a female, as it did not give a croak. Then, the survival chance at the single croak was not $1/2$, but trivially $P_\text{single}(f\vert \bar{\text{c}})=1$, where $\bar{\text{c}}$ denotes the absence of a croak. Therefore, the authors of Ref.~\cite{youtube} apparently solve their riddle  in an inconsistent way, as they do not apply the mentioned rule to the single frog.
These problems arise, because TED wants to pass to us the information 'at least one of the two frogs is male', or in other words 'not ff', by introducing an ill-suited croaking mechanism. As pointed out already in the comments of Ref.~\cite{youtube}, an alternative rule, such as for instance 'if there are two females next to each other, they start fighting', was sufficient to avoid such difficulties.

%\paragraph{Content this work:} 
We here try to solve the frog riddle in its original description, in a logically consistent way, with as few as possible extra assumptions. 
To do so, we define a probability for a male frog to remain silent during the short time of our observation. Eventually, we can study the chance of survival in the two limits of very rare or very frequent croaking, in order to estimate a conclusive answer.  
 Treating the frog riddle this way, we review some of the fundamental aspects of probability theory, that challenge intuition: Are the sexes of the two frogs independent after a croak? Is the labeling important? How is the information 'no ff' received, and do the events of the initial sample space remain equally likely after obtaining this information? 
In the context of these questions, we further demonstrate how to map the frog riddle onto the classic boy-girl paradox. 

\section{Hidden correlations}
First of all, let us discuss why an intuitive assumption of independent frogs, which simply leads to a survival chance of $1/2$ for the two frogs, is not applicable in the riddle. To visualize this point, let us assume for the moment, that we knew which of the two frogs croaked (as we see below, this in fact does not change the odds). 

If all frogs remained uncorrelated after the croak, and, for example, the left frog croaked, we could rule out the two events ff and fm, such that we are left with mm and nf. Given that these two samples are equiprobable, we end up with a chance of $1/2$, that the right frog is female. Equivalently, we could argue, that the frog that
remains silent, is uncorrelated to the croaking one, such that its sex should be male or female with equal chances, i.e. $1/2$. Both these arguments are \emph{not} applicable for the given problem. This becomes apparent, when comparing the above chance with the discussion of the previous section: Of two poisoned persons in the jungle, one being blindfolded and the other one able to see the croaking frog, the 
seeing person would end up with a \emph{different} chance of survival ($1/2 < 2/3$) than the other person. However, as both persons, by rule, catch both frogs, identifying the croaking frog cannot change the odds.

\begin{figure}
\centering
\includegraphics[width=0.99\linewidth]{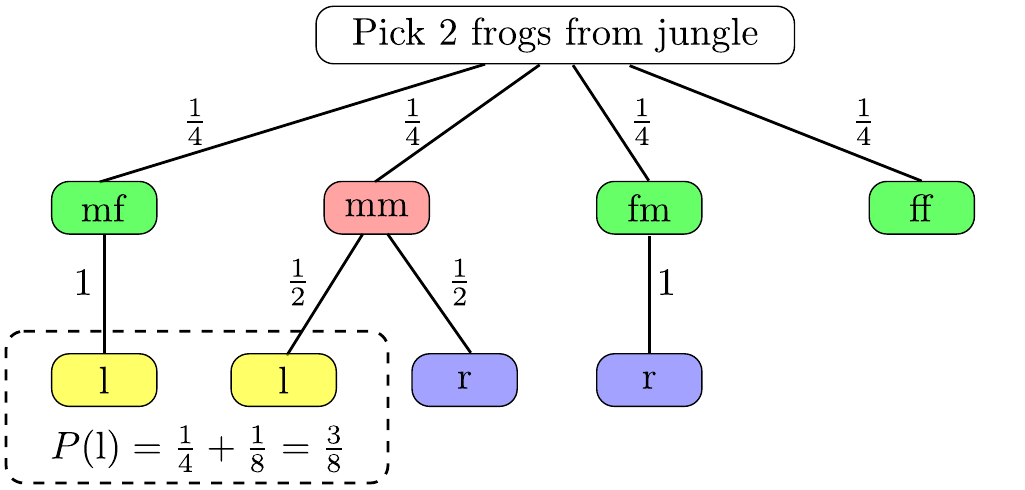}
\caption{ (color online) Probability tree for a scenario where the croaking frog was spotted at the left hand side (l, yellow). Again, the chance to find a female frog (f, green) evaluates to $2/3$. Importantly, seeing the left frog croak is twice as likely in mf as in mm, because in mm the right frog could have croaked as well.} 
\label{fig:seeingthefrog}
\end{figure}

The origin of this alleged paradox is, that the two frogs are no longer independent \emph{after} the croak. In fact, the two remaining events mf and mm (given that we saw the left frog croak), are not equally likely \emph{after} the croak. 
Here, our intuition is misguided by the peculiar croaking mechanism in the riddle. The authors generally allow only for one single croak, meaning that in case of two males mm, both could have croaked with an equal chance. Given that we saw the left (l = '\textit{left}') frog croak, we find the survival chance
\begin{equation}
P(\text{mf}\vert \text{l}) = \frac{P(\text{l}\vert \text{mf}) P(\text{mf})}{P(\text{l})}=\frac{1 \cdot 1/4}{3/8}=2/3.
\end{equation}
 The probability tree in Fig.~\ref{fig:seeingthefrog} describes this situation and gives the probabilities
  used in the above equation. Once the left frog revealed its sex by a croak, the sex of the right frog is no longer independent, and the second (right) frog is more likely to be female than male, because if it was male, it could have croaked instead.

Let us give a brief example of when the argument of independent frogs would be correct. Imagine, that male frogs could only be stimulated to croak, if they get tickled. In that case, if we choose one of the two frogs (in an arbitrary way) and tickle it, and subsequently it produces a croak, we know that it is male. The other frog, however, is still completely independent of the first frog, as there was never a chance for it to give a croak. As a consequence, its chance to be female is simply $1/2$ instead of $2/3$.

Finally, we argue by a rearrangement of the sample space, that it does not change the odds, if we identify which of the two frogs has croaked. This is based on the fact that, by definition, we are allowed to catch both the frogs in the clearing in order to find the antidote. Crucially, we need to be careful in comparing the respective sample spaces after the croak. In case we can not see the croaking frogs, we use the three equally likely events mm, mf, and fm (see first section). Seeing for instance the left frog croak, we conclude, that only the two events mm and mf are possible, however, with non-equal probabilities (see the above calculation). 
This apparent discrepancy between the two approaches is only due to the fact, that in the above labeling, our mind associates the first letter with the left frog. Using a different labeling, we can map both scenarios onto the same, rearranged sample space.
To do so, let us now define for the moment, that the first letter of the labeling (write it uppercase) represents the frog we \emph{heard croaking}, no matter whether or not we know, if it sits to the left or right hand side. As only male frogs can croak, the samle space is for \emph{both} cases $\{\text{Mf}, \text{Mm}\}$. Again, Mf and Mm are no  equally likely events, and the chance to find a female is $2/3$, irrelevant of whether we saw the croaking frog or not.

\section{Solution of the frog riddle}
We now tackle the frog riddle without making use of the implicit rule 'male frogs always croak, but if there are two males, only one croaks', that leads us to logical inconsistencies, as we have seen above.
Instead, we consider the following reasoning a natural interpretation of the narrative of the riddle. 
Each male frog croaks individually, and hence, in case of two males, more than one croak is allowed. 
In fact, let us denote the chance for each male frog to croak $k$ times by $p_k$. 
Explicitly, $p_0$ is the probability for a male frog to remain silent, and $p_1$ the chance to croak exactly once.

\begin{figure}
\centering
\includegraphics[width=0.99\linewidth]{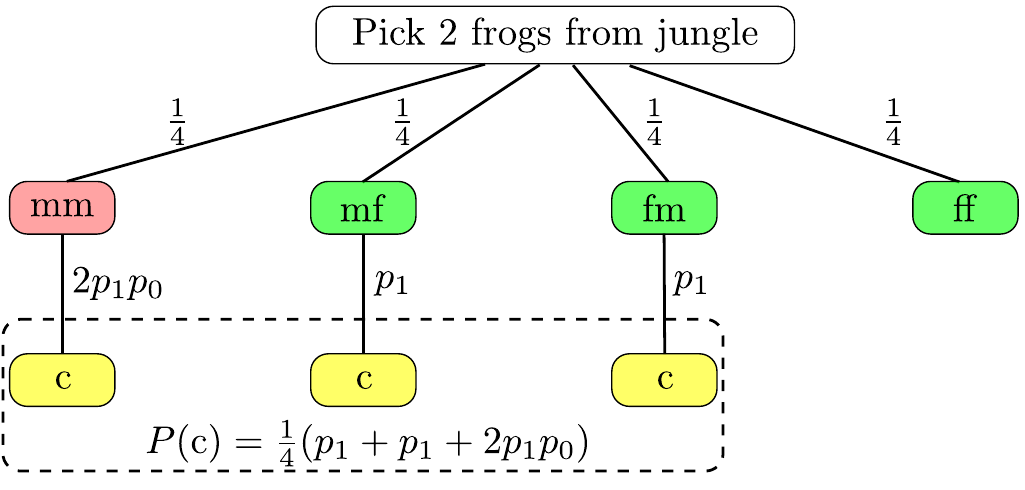}
\caption{ Probability tree depicting a croaking mechanism, where all male frogs croak individually. For simplicity, we show only branches which result in a single croak c. A male frog m croaks with probability $p_1$ exactly one time, and remains silent with probability $p_0$. For the sample mm, one frog has to croak while the other must remain silent, in order to produce exactly one croak.}
\label{fig:solution}
\end{figure}

According to the riddle, we now hear exactly one croak, coming from the clearing with two frogs. 
First, we calculate the chance of the event c = '\textit{exactly one croak}'. 
While for the sample ff, no croak is possible, both mf and fm produce a croak with a chance $p_1$. 
Two males, however, croak exactly one time with the probability $2p_1p_0$, where the factor of two arises because it does not matter  which of the two frogs croaks and which remains silent. 
With that, we find 
\begin{equation}
P(\text{c}) = 1/4\cdot (p_1 + p_1 + 2p_1p_0) = 1/4 \cdot 2 p_1 (1+p_0).
\end{equation}
The survival chance, which is the probability of finding a female frog, therefore is
\begin{align}
P(\text{f}|\text{c}) &= 1-P(\text{mm}|\text{c}) = 1-\frac{P(\text{mm})P(\text{c}|\text{mm}) }{P(\text{c})} \nonumber\\ &= 1- \frac{1/4 \cdot 2 p_1 p_0}{1/4 \cdot 2p_1(1+p_0)} = \frac{1}{1+p_0}.
\label{eq:solution1}
\end{align}
The corresponding probabilities are illustrated in Fig.~\ref{fig:solution}.
Let us briefly interpret this result for possible values of $0<p_0<1$. If a male frog is likely to remain silent, $p_0\to 1$, the survival chance approaches $1/2$. This becomes more intuitive, if we realize, that in this limit male and female frogs almost behave identically. As the second frog did not croak, and thus, we did not receive information about it, it could be female or male with equal chances. 
On the other hand, if croaking is a very likely event, $p_0\to 0$, the survival chance approaches one. This makes sense, because then we have to conclude, that one of the two frogs is female - it would have croaked otherwise. 

The survival chance for the two frogs is always greater or equal to $1/2$. Should we therefore forget about the single frog, and catch the two frogs? Not necessarily, because as the single frog did not croak, it also passed information to us. Our survival chance at the single frog, given that it did not croak ($\bar{\text{c}}$ = '\textit{no croak}'), is
\begin{align}
P_\text{single}(\text{f}\vert\bar{\text{c}}) &=  \frac{P_\text{single}(\bar{\text{c}}\vert \text{f}) P_\text{single}(\text{f})}{P_\text{single}(\bar{\text{c}})}\nonumber\\
&= \frac{ 1 \cdot 1/2}{1/2(1 + p_0)} = \frac{1}{1+p_0}.
\end{align}
Therefore, we find the \textit{same} survival chance in either direction. Clearly, the single frog does not necessarily have a chance of $1/2$ to be female: As croaking becomes more likely ($p_0$ decreases), we have to be more skeptical about whether the frog is really a male, if it remains mute. 

Why does the survival chance actually depend only on $p_0$ but not on the chances of croaking one or multiple times? 
This is because the croaking frog already identified itself as a male. 
It does not matter, how likely it would have croaked a second or a third time, we have already obtained the information that this frog does not provide the antidote. 
In fact, the important information we receive is not the croaking event, but the absence of a croak from the second, or the single frog. 
As we search a female frog, and females do not croak, the key parameter is $p_0$, which is the chance that a frog remains silent although it is male. Note that it
is therefore not important, what kind of statistical distribution (such as a Poisson distribution) we assume to model multiple croaking -- the explicit chances $p_k$ with $k\geq 1$ do not enter into the calculation. 

The finding, that the chances are the same for the two frogs in the clearing, and for the single frog on the tree trunk, does not rely on the introduction of the above parameters $p_k$, but holds generally as long as each frog behaves independent from the others. To see this, let us handle the three frogs in a common sample space. Similar to before, let us now refer with the first (uppercase) letter to the sex of the frog we heard croaking, with the second letter to the other frog in the clearing (that remained silent), and with the last letter to the single frog on the tree stump. We can then write the full sample space as $\{ \text{Mmm}, \text{Mmf}, \text{Mfm}, \text{Mff}\}$. 
The frog that croaked is identified as a male M. For the two remaining frogs, both sexes are possible. Note that these four events are not all equally likely.  However, as each frog croaks individually, the elements Mmf and Mfm are equiprobable (3 frogs, of which 2 are male, produce one croak). 
Therefore, the presented sample space is symmetric under permutation of the second and the third letter. 
From this observation we infer, that even without the introduction of chances $p_k$, the survival chance must be the same in both directions. This holds \emph{generally}, 
as long as all the frogs are treated on equal footing, and we are allowed to catch both frogs on the clearing. Note that the former was not given in the solution presented by TED.

We therefore suggest, that the solution to the riddle -- which way should you go? -- should be phrased the following way: It does not matter -- in both directions there is the same chance of survival. The actual probability of finding an antidote is given by Eq.~(\ref{eq:solution1}), and depends on how likely it is, that male frogs pretend to be female by remaining silent. 
One could make a point for the assmumption that hearing a croak is quite unlikely in a very short observation time of the poisoned person. In this limit, $p_0\to 1$, the survival chances in either direction are about $1/2$. Interestingly, if we assume (somewhat arbitrarily) that $p_0 = 1/2$, we obtain the same value as the authors of Ref.~\cite{youtube}, which is $2/3$ -- however, with a completely different approach and in both directions (tree stump and clearing).

\section{The Boy-Girl paradox}
For problems like the above frog riddle, intuitive thinking usually fails to recognize statistical correlations, that are 
associated with conditional probabilities.
This point has already been convincingly demonstrated by M.~Gardner, who originally presented the "boy-girl paradox" in the ``Scientific American'' \cite{Gardner54}. In a simple form, it illustrates the importance of the way information is obtained. Up to now, the boy-girl problem has been subject to numerous reformulations and reinterpretations (e.g. \cite{Carlton05}). As we explain below, the frog riddle can be viewed as a modified version of the latter.

The original paradox has two main variations:
\begin{itemize}
 \item Mr. Smith has two children. The older child is a boy. What are the odds that both children are boys?
 \item Mr. Smith has two children. At least one of them is a boy. What are the odds that both children are boys?
\end{itemize}
Throughout, we assume, that the probability for any individual child to be a boy or a girl is about $1/2$.
The answer to the first question then is simply $1/2$, since for the younger child there is an equal chance of being male or female. Because of the distinction ``the older'' and ``the younger'', 
we have a knob to discriminate between the two children, and both become independent of each other.
The answer to the second question is more tricky and, as it turns out \cite{Gardner61, BarHillelFalk82}, depends on how the information ``at least one of them is a boy'' 
is actually obtained. 

Let us first assume that we meet Mr. Smith, and he tells us, that at least one of his two children was a boy. In this case, the children remain indistinguishable to us. 
Our original sample space (before Mr. Smith talks to us) was $\{$BB, BG, GB, GG$\}$, all being equally likely with a chance of $1/4$. Learning that one of the children is a boy, rules out the sample GG, 
such that we remain with the equally likely possibilities $\{$BB, BG, GB$\}$, and the chance of two boys is $1/3$. 
This is exactly the intended version of the frog riddle, leading to the $2/3$ chance of a girl (or female frog), see Eq.~(\ref{eq:intended}) and Fig.~(\ref{fig:tree0}). Importantly, no logical inconsistencies 
can occur here, in contrast to what happens in the frog riddle (see above).

Next, we modify the scenario: We meet Mr.~Smith on the street with a boy, that he introduces as his son. In this case, the chance for a second son is
\begin{equation}
 P(\text{BB}|\text{B}_\text{m})=\frac{P(\text{B}_\text{m}|\text{BB})\, P(\text{BB})}{P(\text{B}_\text{m})}= \frac{1\cdot1/4}{1/2}=1/2.
\end{equation}
Here, the index m denotes the event of meeting a child, and $P(\text{B}_\text{m})=1/2$ is the chance that Mr.~Smith chose his boy for his walk. Our odds have now increased from $1/3$ to $1/2$. How is that possible since we seemingly obtained the same information?
Let us consider the diagram \ref{fig:walk}:
\begin{figure}[]
\centering
\includegraphics[width=0.99\linewidth]{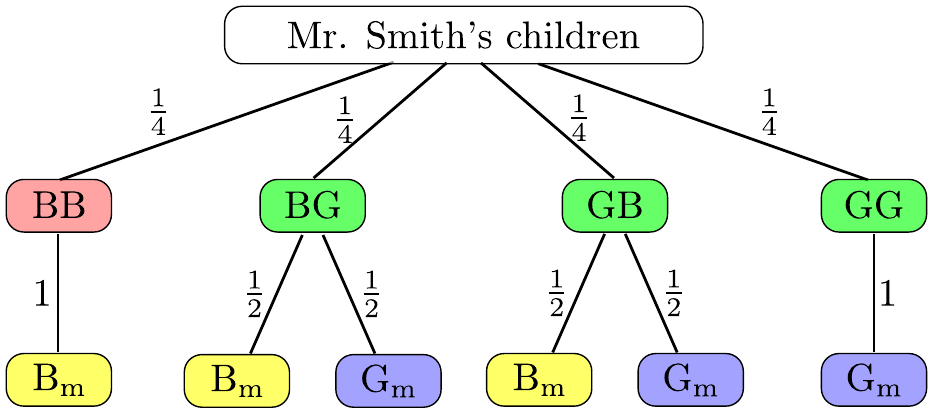}
\caption{(color online) Probability tree for a scenario, where Mr. Smith randomly picks one of his two children for a walk. If a boy is chosen ($\text{B}_\text{m}$, yellow), the chance for the second child to be a girl (G, green) is $1/2$.
} \label{fig:walk}
\end{figure}
As we can see, the chance of meeting a boy, given that Mr. Smith has a boy and a girl, is $1/2$, assuming that Mr. Smith \textit{randomly} picks one of his children for the walk. As the gender of the child does not affect the choice of Mr.~Smith, one can argue by means of independent samples: As the gender of the second child is independent of the gender of the child we met, the chances for a boy (or girl) for the second child are $1/2$.

A different way of approaching the problem is to note, that the sample space after we met a boy, which is $\{\text{BB}, \text{BG}, \text{GB}\}$, does not consist of equally likely events. The event BB is as likely as BG and GB \emph{together}, after we met the boy, cf.~Fig.~\ref{fig:walk}. 
The difference to the former situation is, that the two children can be labeled by m =  '\textit{child
we met}' and h = '\textit{child at home}'. 
With these indices, the sample space can be written in terms of equally likely events, $\{ \text{B}_\text{m} \text{B}_\text{h},$ $\text{B}_\text{h} \text{B}_\text{m},$ $\text{B}_\text{m} \text{G}_\text{h},$ $\text{B}_\text{h} \text{G}_\text{m},$ $\text{G}_\text{m} \text{B}_\text{h},$ $\text{G}_\text{h} \text{B}_\text{m},$ $\text{G}_\text{m} \text{G}_\text{h},$ $\text{G}_\text{h} \text{G}_\text{m}\}$. 
With this notation, we effectively compensated for the greater probability of $P(\text{B}_\text{m}|\text{BB})$, by explicitly writing  $\text{B}_\text{m} \text{B}_\text{h}$ and $\text{B}_\text{h} \text{B}_\text{m}$.  
The condition that we met a boy, reduces the sample space to $\{ \text{B}_\text{m} \text{B}_\text{h}, \text{B}_\text{h} \text{B}_\text{m}, \text{B}_\text{m} \text{G}_\text{h}, \text{G}_\text{h} \text{B}_\text{m}\}$.
Two of the four samples of the sample space contain two boys, leaving us with a chance of $1/2$, that the child at home is a boy as well.
Indeed, \textit{any} situation that allows us to differentiate between the two children \emph{in a meaningful way}, will change the sample space, and thus the odds. For the same reason, for instance, a little boy who is about to get a baby sibling, can not predict the gender of the baby from fact that he is a boy \cite{BarHillelFalk82, Carlton05}. The other baby has a fifty percent chance of being a brother or a sister, since the two children (born and unborn) are clearly independent, and hence, distinguishable.

Meeting the boy in the street is therefore very similar 
to the statement ``the older child is a boy'', since we can simply replace the label ``older'' by ``in the street''.
This analogy gets even more clear when we notice that the probability remains $1/2$, if we meet Mr.~Smith with a boy that
he introduces as his oldest child. In this case ``the oldest child'' does not provide us with relevant further information, since we have already labeled the two children by the encounter of one 
child in the street \cite{BarHillelFalk82}. Importantly though, the labeling is not performed by us seeing the son, but by Mr. Smith randomly picking one child for the walk.
This is analogous to the previously mentioned case of randomly selecting one frog, and subsequently stimulating it to croak (for instance by tickling).

As we already discussed in the previous section, the sexes of the frogs are not independent in the intended version of the frog riddle in Ref.~\cite{youtube}. This is because the authors there use the assumption, that only one croak is allowed at maximum, and as a consequence, a possible second croak from a second male is suppressed. Therefore, even though we can distinguish the two frogs by the label 'the one that croaked' and 'the one that remained silent', the authors find a chance for a female frog that is different from $1/2$.

Let us emphasize the importance of the selection mechanism with the following example. Imagine that Mr.~Smith belongs to a culture where girls never join for walks with their parents. With such a biased selection mechanism, seeing Mr.~Smith with a son then does not give us any information beyond the fact that he has 'at least one boy'. The chance for the other child to be a girl is therefore again $1/3$. Why does the previous analysis with a distinction of the child at home and the child on the walk not apply here? In both cases, $\text{BG}$ and $\text{GB}$, the labels m = 'met' and h = 'home' are meaningless, as the gender already determines which child will accompany Mr.~Smith. Distinguishing $\text{BB}$ in the form of $\text{B}_\text{m}\text{B}_\text{h}$ and $\text{B}_\text{h}\text{B}_\text{m}$ is in principle possible, but does not result in a sample space of equally likely events. This sample space is in fact given by $\{ \text{BB}, \text{BG}, \text{GB}\}$, in which one out of three contains a second boy. This scenario also corresponds to the intended version of the frog riddle, see Fig.~\ref{fig:tree0}. There, one male is randomly selected (but never a female) and reveals its sex by a croak. The female frogs never could have revealed themselves, just like here we can not meet Mr.~Smith's girls on the street.
%%%%%

The above analysis illustrates how the way information is received affects the conditional probabilities of the boy-girl problem. 
It is furthermore insightful to investigate, how additional information, for instance about further attributes of Mr.~Smith's children, alters the odds. 
This is shown to correspond to the frog riddle under the assumption, that male frogs croak at a given chance. Similar to the probability of a croak in the latter, the chance of occurrence of the attribute is essential in the boy-girl paradox.
An example, that has been discussed exhaustively in the literature \cite{Lynch11, Falk11}, 
is the one of Mr.~Smith (of whom we know by now that he has two children) letting us know, that he has at least one boy who was born 
on a Tuesday. To make a connection to the frog riddle, let us 
here study a slightly modified version of this scenario.
Assume that we meet  Mr. Smith with a boy in the street. At this point, since we can label the boy by 'the boy we met', and under the assumption that Mr.~Smith randomly selects a child  to go with him for a walk, there is a chance of $1/2$ that the other child (at home) is a girl (see above). However, as we greet Mr.~Smith, the boy happens to inform us: 'I am the only boy in my family who was born on a Tuesday'. How have the odds changed, now that we have obtained this information? Importantly, the two attributes have to be considered a logical conjunction -- the other child can still be a boy, or could be born on Tuesday, but not both. 

We denote the chance for being born on a Tuesday by $p_1 = 1/7$, and to be born on any other day of the week by $p_0 = 1-p_1$. The probability that Mr. Smith has another daughter then is again given by Eq.~(\ref{eq:solution1}), so it reads $P(\text{G}|\text{c})=1/(1+p_0)$, where compared to the frog riddle, the croaking event c is replaced by the event c = '\textit{boy met who was born on a Tuesday}'. 
Explicitly, the chance for a second girl yields approximately $P(\text{G}|\text{c})=0.54$. The effect on the probability becomes more intuitive in following limits: Had the boy said, that he was the only child that is male and born on new years eve, $p_1 = 1/365$, we would end up with a chance of $P(\text{G}|\text{c})=0.5007$  to find a girl. As being born on new years eve is a very rare event, the information that he is the only child with this feature is almost not affecting the odds. For the other child, both the options of a boy \emph{not} born on new years eve, and of a girl born on any day, are almost equally likely. On the other hand, if the boy tells us that he is the only child in the family that is male and born \emph{not} on new years eve, $p_1 = 364/365$, the probability to find a girl evaluates to $P(\text{G}|\text{c})=0.997$. Because not being born on new years eve is in turn a very common attribute, it is very likely that the other child is a girl -- after all, it needs to comply with one of the two constraint 'female born on any day' or 'boy being born on new years eve'.

\section{Conclusion}

We have reviewed how information, and the way information is obtained (events that transpire in different ways), can affect the probabilities of samples in the sample space. 
In many cases, this makes the assessment of chances challenging for our intuition, 
as we generally tend to disregard statistical correlations between different outcomes.
In both the frog riddle and the boy-girl problem, for example, the actual 
mechanism of transferring information is essential.
The frog riddle was drafted as a modified version of the original boy-girl paradox, however, with a subtly altered way of conveying information. 
As we show in this article, the solution as presented in Ref.~\cite{youtube}, does not account for this point, and instead, different rules are applied to different parts of the game (for instance, a potential croak from the single frog is not considered).

Here, we present a modified solution to overcome this problem, finding, that the chance of survival varies between $1/2$ and $1$, depending on the probability of male frogs to croak. Importantly, the same odds are obtained in both directions (single or two frogs), such that any of the two can be chosen, which is in contrast to the strategy suggested in Ref.~\cite{youtube}.
The possibility of multiple croaking, and its statistical distribution, does not affect the result.

We thank F. Keidel and C. De Beule for stimulating discussions, as well as B. Mayer for bringing the frog riddle to our attention.


\begin{thebibliography}{3}


\bibitem[Abbott (2016)]{youtube}
D. Abbott (TED-Ed), {\url{https://www.youtube.com/watch?v=cpwSGsb-rTs}} and comments, as well as reply videos. Date 02/29/2016.
\bibitem[Gardner (1954)]{Gardner54}
M. Gardner (1954), The Second Scientific American Book of Mathematical Puzzles and Diversions. Simon \& Schuster. ISBN 978-0-226-28253-4
\bibitem[Gardner (1961)]{Gardner61}
M. Gardner (1961). The Second Scientific American Book of Mathematical Puzzles and Diversions. Simon \& Schuster. ISBN 978-0-226-28253-4.
\bibitem[Bar-Hillel \& Falk (1982)]{BarHillelFalk82} 
M. Bar-Hillel and R. Falk (1982), ``Some teasers concerning conditional probabilities.'', Cognition 11 (2), 
doi:10.1016/0010-0277(82)90021-X
\bibitem[Lynch (2011)]{Lynch11}
P. Lynch (2011), Irish Math. Soc. Bulletin 67, 67-73.
\bibitem[Falk (2011)]{Falk11}
R. Falk (2011), ``When truisms clash: Coping with a counterintuitive problem concerning the notorious two-child family''. Thinking \& Reasoning, Vol. 17, issue 4, p.353-366,
doi:10.1080/13546783.2011.613690
%\bibitem{wiki}  \url{https://en.wikipedia.org/wiki/Boy_or_Girl_paradox}.
\bibitem[Carlton \& Stansfield (2005)]{Carlton05}
M. A. Carlton and W. D. Stansfield (2005). "Making Babies by the Flip of a Coin?". The American Statistician. 59, p. 180–182, doi:10.1198/000313005x42813.
 
\end{thebibliography}
\end{document}